# Dirac energy spectrum and inverted band gap in metamorphic InAsSb/InSb superlattices

Sergey Suchalkin[1*], Maksim Ermolaev[1], Tonica Valla[2], Gela Kipshidze[1], Dmitry Smirnov[3], Seongphill Moon[3,4], Mykhaylo Ozerov[3], Zhigang Jiang[5], Yuxuan Jiang[5], Stefan P. Svensson[6], Wendy L. Sarney[6] and Gregory Belenky[1],

[1]*Stony Brook University, Stony Brook, New York, 11794, USA*

[2]*Brookhaven National Laboratory, Upton, New York, 11973, USA*

[3]*National High Magnetic Field Laboratory, Tallahassee, Florida,, 32310, USA*

[4] *Department of Physics, Florida State University, Tallahassee, Florida 32306, USA*

[5]*Georgia Institute of Technology, Atlanta, Georgia, 30332, USA*

[6]*U.S. Army Research Laboratory, 2800 Powder Mill Rd, Adelphi, Maryland 20783, USA*

A Dirac-type energy spectrum was demonstrated in gapless ultra-short-period metamorphic InAsSb/InSb superlattices by angle-resolved photoemission spectroscopy (ARPES) measurements. The Fermi velocity value $7.4 \times 10^5$ m/s in a gapless superlattice with a period of 6.2nm is in a good agreement with the results of magneto-absorption experiments. An "inverted" bandgap opens in the center of the Brillouin zone at higher temperatures and in the SL with a larger period. The ARPES data indicate the presence of a surface electron accumulation layer.

The virtual substrate approach[1] makes it possible to fabricate narrow gap InAsSb – based superlattices (SL) with ultra-thin layers, formed by engineered ordering of the group V element composition[2]. Such structures allow easy bandgap control and tailoring of the carrier dispersion[3]. The exceptional design flexibility originates from the large lattice mismatch and broken gap band alignments between the corresponding binaries - InAs and InSb[4]. The electronic and optical properties of $InAs_{1-x}Sb_x/InAs_{1-y}Sb_y$ SL differ from the bulk InAsSb alloys. While the minimum bandgap of the random alloy, reached at the Sb composition ~ 0.6 is ~ 100 meV[5-7], the bandgap of an InAsSb/InSb superlattice can be reduced to zero and beyond to negative values by increasing the SL period[3, 8]. These engineered semiconductor materials can host nontrivial topological states[9, 10]. For example, a new topological semimetal (TSM) phase - a triple point TSM is predicted in an InAsSb alloy with short period group V composition ordering along [111] direction[10, 11]. High intrinsic spin-orbit coupling, high carrier mobility, the possibility of magnetic doping and an ability to form a good interface with a superconductor (Al) makes the $InAs_{1-x}Sb_x/InAs_{1-y}Sb_y$ SL a perfect candidate for a new material platform for quantum information devices.

To probe the bandgaps of alloys and SLs down to ~ 70 meV, photoluminescence can be used. For the nearly gapless SLs, alternative experimental method for the investigations of the band structure such as infrared magneto-absorption must be used[7, 12-14]. More detailed information can be gained through the angle resolved photoemission spectroscopy (ARPES) which is a direct experimental method to study the band structure and carrier dispersion of materials[15] near the surface. However, application of this method to many SLs is limited since the ARPES probing depth (few nm) is usually shorter than the SL structure period. For example, the semiconductor-semimetal transition in a pseudomorphically grown InAs/GaSb SLS on GaSb happens at a period of ~ 17 nm, so using ARPES to obtain electron dispersion in such structures might be problematic. The bandgap of our new metamorphically grown InAsSb/InSb SLs becomes 0 at a SL period of 6.2 nm and ARPES can be used to monitor carrier dispersion near the semiconductor-semimetal transition point.

In this work we present ARPES studies of nearly gapless short-period InAsSb/InSb metamorphic superlattices. The ARPES setup includes a Scienta R4000-WAL electron spectrometer with a wide acceptance angle (~30º), providing for simultaneous detection over a wide range of the momentum space with ~ 0.1º precision[16,17]. The photon source is a Scienta VUV5k microwave-driven plasma-discharge lamp with a monochromator, providing a He *I* radiation (21.22eV). The combined energy resolution used in the present study was 5 meV. A fully motorized 5-axis manipulator is equipped with a Janis flow cryostat which allows cooling the sample down to 6 K. Relatively low resolution with respect to the momentum component $k_Z$ along the SL growth direction resulted in the averaging of the ARPES spectra over the entire range of $k_Z$ in the SL Brillouin zone.

The structures grown for ARPES are InAsSb/InSb superlattices with periods of 6.2 nm and 7.8 nm. The structure details are identical to those reported in ref. [8], except that the ARPES samples were grown without either a top barrier or a cap layer. The SLs used for the ARPES studies were terminated with the ternary layer. The sample surfaces were protected from contamination by low-temperature deposition of a thick As layer in the MBE chamber.

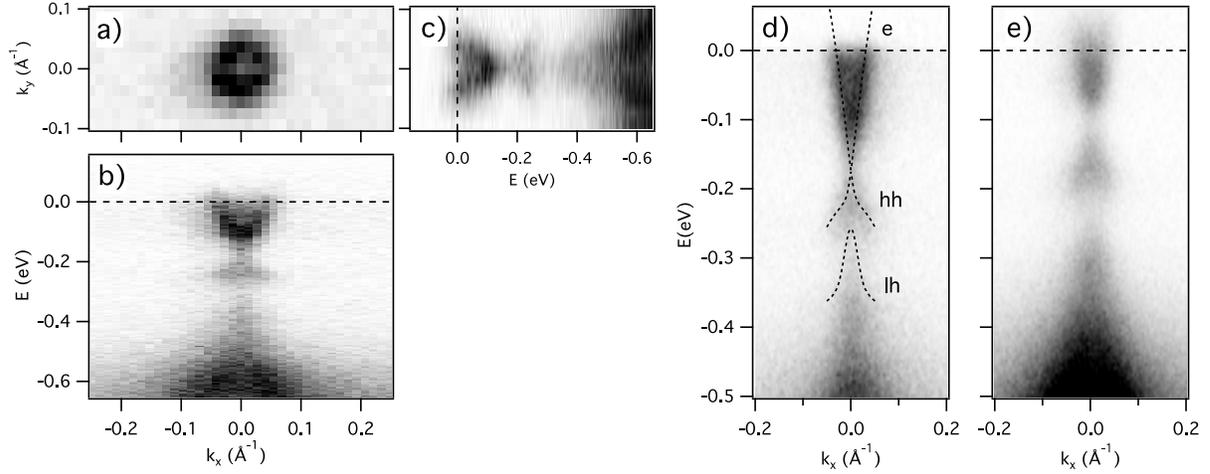

Fig. 1. The electronic structure of InAs$_{0.48}$Sb$_{0.52}$/InSb SL with a period of 6.2 nm. The Fermi surface (a), and dispersion of the low-energy excitations along the two perpendicular momentum lines (b and c) indicate an isotropic in-plane electronic structure. The low-temperature electronic structure (a-d) is nearly gapless. The 300 K spectrum (e) shows a finite gap. Dashed lines in (d) are band-fits based on 8x8 k*p calculations[18].

After the MBE growth, the wafers were transported to the ARPES setup in Ar-filled containers and then loaded into the analysis chamber where the protective cap was removed by heating the sample to ~300 C. The effectiveness of the As cap was checked at both facilities: we exposed a sample to air for 24 hours, then placed it back to the MBE chamber, annealed off the cap and verified that a good REED pattern could again be observed, while in the ARPES chamber, both the low-energy electron diffraction (LEED) and ARPES indicated a high quality of de-capped surfaces.

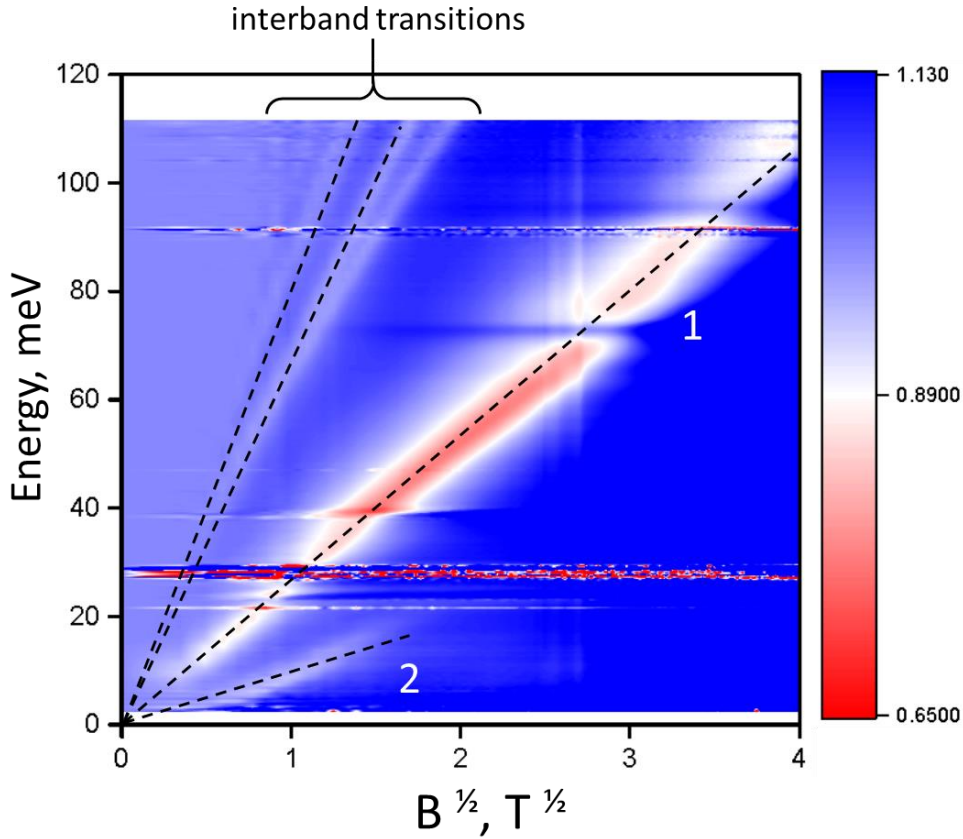

Figure 2. Color plot of the absorption as function of the emission energy and square root of the magnetic field. Line 1 corresponds to the transition between the 0th and 1$^{st}$ electron Landau levels, line 2 – to the transition between the 1$^{st}$ and 2$^{nd}$ electron Landau levels. The dashed lines are guides to the eye.

The ARPES spectra for two temperatures (40 K and 300 K) of the SL with a period of 6.2 nm are presented in Figure 1. Signal intensity is shown in the gray-scale where the darker corresponds to higher intensity. At low temperature, the in-plane low-energy electronic structure shows an isotropic "hourglass" like dispersion near the zone center (Fig. 1(a-d)) with the conduction and valence bands practically touching, indicating a nearly gapless SL. A limited energy resolution and an intrinsic broadening of the states does not allow us to conclude whether this is a truly gapless SL or a small gap of about a few meV is present. The zero gap is obtained from 8 band k*p calculations, as shown in Fig. 1(d), indicating that this SL is a 3D Dirac semimetal.

The increase in temperature leads to the band energies shifting to higher values and the opening of a sizable gap (Fig. 1,(e)), approximately 40 meV. Based on the sign of the experimental temperature coefficient of InAsSb bandgap[19] and the k*p calculations with temperature dependent-parameters, we attribute the observed bandgap to the hybridization effects where the conduction and valence bands overlap, producing and inverted band-gap, with the SL becoming semi-metallic.

The Fermi energy extracted from the low temperature ARPES data is ~150 meV, roughly 4 times higher than that obtained from the bulk Hall measurements[20]. The most likely reason for this is the presence of a surface accumulation layer[21-23]. The Fermi energy value is in reasonable agreement with estimates based on knowledge of the position of the Fermi-stabilization level and the bending of the conduction band edge energy near the surface. Additionally, nonequilibrium carriers induced by the probe illumination may contribute to an excess band filling. The signal from the SL valence band is weaker than that of the conduction band since the holes are mostly localized in the InSb layers and the first hole containing InSb layer is located deeper under sample surface. The value of the Fermi velocity near the Γ point obtained from a 8x8 k*p model (Fig 1, a) is $7.4 \times 10^5$ m/s [8]. This is somewhat higher than that obtained from our previous ternary-ternary SLs [3]. The difference can be explained by better electron-hole overlap in the InAsSb/InSb SL.

The Dirac character of the carrier dispersion, directly observable in ARPES studies, is also confirmed by the results of the magneto-absorption measurements (Fig 2). The square root dependence of the cyclotron resonance transition energy (lines 1, 2 in Fig. 2) on the magnetic field is a manifestation of the linear dispersion[24]. The gapless character of the SL is established by the zero intercept of the concurrent lines corresponding to the interband transitions.

Based on the calculations, a hybridization gap which appears at 300K results from the band overlap and anti-crossing, so the SL becomes semi-metallic. Characteristic features of the carrier dispersion in the semi-metallic SL having cubic crystal symmetry are: a gap formed by band anti-crossing in the in-plane direction at zero wave vector component in the growth direction ($k_Z=0$) and two anisotropic Dirac points due to band crossing at two symmetrical points $\pm k_{Z0} \neq 0$ (Fig. 3,b)[22, 25-27]. Such an energy spectrum is, in general, similar to that predicted in InAsSb alloys with CuPt type ordering[10]. In our case the growth direction is [001], so the SL has the symmetry of an InAsSb alloy with CuAu type of ordering[28]. Taking into account the lack of the inversion symmetry in zinc blend crystals and the fact that the order of [001] symmetry axis ($S_4$) is different from this of [111] ($C_3$) we can expect that the "fine structure" of the crossing points in our case will be different from that described in[10] . The detailed analysis of the band degeneracies at the crossing points will be given elsewhere. The nanoscale ordering of the SL reduces the Td symmetry thus allowing linear in k, spin orbit splitting terms (Rashba terms)[29], making the SL a new material for realization of "triple point" topological semimetals[10].

The transition in the electronic structure with temperature observed here for a nearly gapless SL is likely the first example of a material that changes its character from a trivial to topological semimetal with a relatively small change in temperature. We expect that a moderate pressure could induce similar transition for a nearly gapless SL.

Increase of the SL period leads to overlap between conduction and valence band of the SL. The ARPES spectrum of the SL with a period of 7.8 nm is presented in Fig. 3,a. In contrast to Fig.

2,(a) a finite gap of approximately 30 meV separating upper and lower parts of the hourglass-like pattern is clearly seen. There is a residual signal in the gap, however, which can be a signature of the band crossings taking place at two symmetrical points with zero $k_{x,y}$ and nonzero $\pm k_{z0} \neq 0$. The ARPES spectra represent the effective average over different values of $k_z$ within the SL Brillouin zone. Since a typical ARPES probing depth with 21eV photons is of the order of 1nm, the averaging is over the whole SL Brillouin zone. Therefore, even in the case of the band gap opening near $k_z=0$, we can expect to see some signal contribution from two Dirac points at some $\pm k_{z0} \neq 0$. The low intensity of this signal can be caused by strong reduction of the density of states as $k_z$ approaches the middle of the SL Brillouin zone.

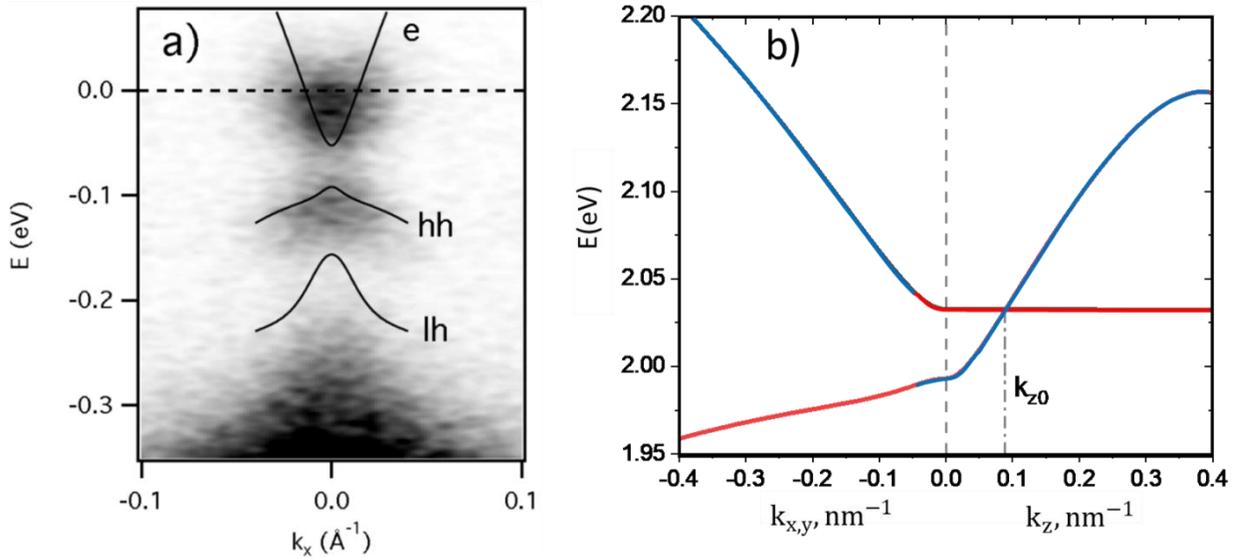

Figure 3. (a) ARPES spectrum of InAsSb/InSb SL with a period of 7.8 nm, recorded at T=108 K. The solid lines represent the 8-band k*p calculations; (b) calculated carrier dispersion of the band-inverted SL for $k_{x,y}$ and $k_z$. Blue curve indicates electron-like states while the red curve indicates the hole-like ones.

In conclusion we have presented measurements of the carrier dispersion in short-period metamorphic InAsSb/InSb ultra-narrow gap superlattices. To our knowledge this is the first report of ARPES probing of SL carrier dispersion. The Dirac character of the dispersion with a Fermi velocity of $7.4 \times 10^5$ m/s in the gapless SL was demonstrated. The energy gap at $k_z=0$ changes from 0 to an inverted one of 30 meV as the SL period is increased from 6.2 to 7.8 nm. The data demonstrate that the character of the energy dispersion in a short period metamorphic InAsSb/InSb SL is similar to the type of dispersion in corresponding bulk alloys with compositional ordering. In contrast to the latter, the SL energy spectrum can be predicted by and controlled with the structure's design. This control makes metamorphic InAsSb/InSb SLs a novel flexible platform for realizing quantum materials and developing a variety of optoelectronic and quantum devices.

Acknowledgments:

The work of M.E., S.S., G.B. and G.K. is supported by National Science Foundation grant no. DMR-1809708, U.S. Army Research Office grant no. W911TA-16-2-0053.and by Center of Semiconductor Materials and Device Modeling. ARPES experiments at Brookhaven National Laboratory were supported by the US Department of Energy, Office of Basic Energy Sciences, contract no. DE-AC02-98CH10886. S.M. and D.S. acknowledge the support from the U.S. Department of Energy (grant no. DE- FG02-07ER46451) for IR magneto-spectroscopy measurements that were performed at the National High Magnetic Field Laboratory, which is supported by National Science Foundation (NSF) Cooperative Agreement no. DMR-1157490 and DMR-1644779, and the State of Florida.